Performance Evaluation of LoRa Technology for Rural Connectivity: An Experimental Analysis in Nepal

Atit Pokharel[1], Pratik Sapkota[1], Dilip Sapkota[1], Shashank Dahal[1], Sulav Karki[1], Ram Kaji Budhathoki [1]

[1]Department of Electrical and Electronics Engineering, Kathmandu University, Dhulikhel, Kavre, Nepal

Correspondence should be addressed to Ram Kaji Budhathoki; ramkaji@gmail.com

## Abstract

LoRa technology has garnered significant interest in the Information and Communications Technology (ICT) field in recent years due to its ability to operate at low power while maintaining effective communication. Despite gaining attention, LoRa technology faces challenges in effectively facilitating communication in rural settings due to specific transmission and reception conditions. This research paper provides an in-depth analysis of using a LoRa mesh network that accesses the performance of different LoRa configurations by varying parameters like Bandwidth (BW), Spreading Factor (SF), and Coding Rate (CR). Metrics, like the Received Signal Strength Indicator (RSSI), Signal-Noise Ratio (SNR), and packet loss, are analyzed to check the optimal configurations for LoRa nodes, specifically in the context of rural areas of Nepal. Furthermore, the varying propagation loss concerning the change in physical layer parameters is also discussed. The experimental setup utilizes Arduino Uno and ESP 32 microcontroller boards with LoRa modules to build the transmitter and receiver nodes, which are paired with a self-constructed monopole antenna, showing superior gain compared to commercially available options. This paper also explores the potential of integrating the acquired data with cloud platforms such as ThingSpeak. This integration establishes a strong backbone for the Internet of Things (IoT), which can gather and analyze remote data, providing the capacity for remote access to the data. This paper finally recommends specific values for the examined parameters for the specific case of a particular type of hilly and mountainous terrain in a country like Nepal, keeping in mind the unique trade-offs each one offers, thereby enabling optimal rural wireless communication. The usability of such networks extends to applications where low data rates along with long-range and less power consumption are required, offering a hopeful future for improved rural connectivity and IoT applications.

**Keywords:** LoRa, Received Signal Strength Indicator, Signal to Noise Ratio, Time Division Multiple Access, Internet of Things

## 1. Introduction

LoRa, a patented modulation developed by Semtech, utilizes the Chirp Spread Spectrum (CSS) modulation, promising long-range, low-power consumption and secure data transmission at low data rates [1]. Commonly referred to as "long-range," LoRa can function with public, private, or hybrid networks, achieving a greater reach than conventional cellular networks [2]. It operates on the Industrial, Scientific and Medical (ISM) band frequencies like 433 MHz, 868 MHz and 915 MHz. Its effortless integration with existing infrastructures has made it a valuable tool for developing cost-effective, battery-operated Internet-of-Things (IoT) applications [1].

In recent years, the exploration of LoRa in various wireless sensor networks, mostly in IoT, has been trending due to LoRa's range and power consumption. LoRa has been widely used and researched in the context of LPWAN (Low Power Wide Area Network) [3]. It is increasingly used with cloud-based platforms to create more robust IoT systems. One such platform is ThingSpeak [4], an open-source, cloud-based platform allowing real-time aggregating, visualizing, and analyzing data streams. The data collected from LoRa nodes can be pushed to ThingSpeak [5], thereby creating an accessible data backbone that supports remote monitoring and analysis. This approach adds a new dimension to IoT applications, particularly in rural or remote areas where direct data monitoring might be challenging. The paper explores this integration potential further, demonstrating how a more connected and responsive IoT system can be realized.

Due to the fact that LoRa is a new product, there are questions about its reliability. In wireless link communication, many connection impairments degrade the connection of the link. Impairments include wireless propagation effects, network interference, and thermal noise [6]. The effects of signal propagation also include attenuation with distance (also referred to as path loss) and the reception of multiple copies of the transmitted signal (referred to as multipath fading) [7]. A key characteristic of LPWAN technologies is the ability to fine-tune the physical layer's settings to select a more sensitive and optimal configuration that allows for the most reliable communication depending upon various conditions and its application. This flexibility makes LPWAN technologies appealing to developers of IoT applications requiring long-range communications with relatively low data rates. At the same time, however, the ability to fine-tune physical layer settings requires a thorough understanding of their impact on network performance, especially on communication's reliability and energy efficiency [8, 9]. The experiment and analysis center on metrics such as Received Signal Strength (RSSI), Signal-to-Noise Ratio (SNR), and packet loss altering the physical-layer parameters like Bandwidth (BW), Spreading Factor (SF), and Coding Rate (CR). The main scope of this work is to compare and contrast these metrics to check the optimal LoRa configurations for wireless connection between LoRa nodes, specifically for a hilly geographic terrain like the Kavrepalanchowk district of Nepal.

LoRa can be leveraged to develop cost-effective mesh networks using Time Division Multiple Access (TDMA) [10]. TDMA offers the advantage of precise synchronization and channel separation, significantly reducing network interference, making it a popular choice for LoRa mesh network deployment. The cost-effectiveness of these networks is particularly attractive for IoT applications that require broad coverage with low infrastructure costs. The TDMA method provides an efficient means to use the available bandwidth, permitting numerous nodes to communicate within the same network and enhancing the communication system's scalability and sustainability [10, 11]. This paper, therefore, aims to enhance the optimal functionality of LoRa mesh networks utilizing TDMA in rural settings where a clear Line of Sight (LOS) can be achieved.

## 2. Previous Works

A lot of real-world studies have been done over the past few years to evaluate the performance of LoRa in different conditions and applications. 'A Study of LoRa Low Power and Wide Area Network Technology' [7] analyzes the performance of LoRa technology based on the code rate, spreading factor, and bandwidth parameters. This paper is instrumental in understanding how these parameters influence LoRa's performance, particularly regarding the data rate and time on air. Its exploration of LoRa's adaptability using the spread spectrum technique provides foundational insights for analyzing LoRa configurations in rural settings like the Kavrepalanchowk district of Nepal. The study's context, focusing on urban smart city applications, presents relevant parallels for extending these insights to rural environments. The results highlight that higher spreading factors result in a more extended communication range, and each spreading code corresponds to a specific spreading factor and is considered orthogonal. On the other hand, it overlooks crucial aspects such as the effects of spreading factor and bandwidth on signal strength parameters like RSSI, SNR, and packet loss. In addressing this gap, our research delves into these unexplored areas, comprehensively analyzing LoRa's performance in rural environments.

In 'Evaluation of LoRa Technology in Flooding Prevention Scenarios' [12], the performance of LoRa technology was assessed regarding the received packet ratio, RSSI, SNR, and network latency. The tests conducted in the Tagus estuary and a rural area near Lisbon, Portugal, provided valuable insights into LoRa's real-world performance. The paper's findings, particularly regarding LoRa's robustness and effectiveness in diverse scenarios, are a notable positive aspect. This study, with its focus on technical performance in various environments, including considerations of node placement, provides a useful comparative framework for our research in rural Nepal. However, the study focused on a limited set of spreading factors and did not analyze packet loss and SNR. Moreover, the influence of node placement concerning tidal levels was highlighted, indicating the need for further exploration in diverse environments to ensure the generalizability of the findings.

Our selection of the SX1278 module for LoRa communication was influenced by the insights from 'Performance Evaluation of Low-Cost LoRa Modules in IoT Applications' [13], which offers an in-depth evaluation of the SX1272 and SX1278 modules. This paper not only compared the transmission range and power consumption of these modules but also investigated their operational efficiency in IoT applications. The power variation between the two modules was observed to be around 30% at maximum transmission power, while at lower settings, the difference was negligible, ranging from 14% to 18%. However, it is essential to note that this research primarily focused on the transmission range and power consumption aspects. It did not delve into other critical factors, such as link quality, packet loss and the performance of the LoRaWAN network. This substantial gap is covered in our paper, which examines the performance of the SX1278 module in a different setting.

The insights from 'Performance Analysis of LPWAN Using LoRa Technology for IoT Application' [3] have been pivotal in shaping our approach to understanding LoRa's capabilities in IoT applications, particularly in rural environments. While [3] analyzes LoRa's signal strength (RSSI), packet loss, and SNR and discovers the stability of these parameters with higher spreading factors, it primarily focuses on the indoor IoT application context. In our study, we extend this analysis to a rural setting in Nepal, investigating system performance under lower bandwidth conditions, which was missing in [3]. Furthermore, we also delved into exploring the dynamics of packet loss and the impact of coding rate in a less controlled environment. This extension not only complements the urban-centric findings of [3] but also provides a nuanced understanding of LoRa's performance in the unique topography of Nepal.

Besides these, numerous other experiments evaluate the performance of LoRa in various conditions. However, the experimental setup, equipment characteristics, geographical factors like terrain, obstacles in LOS, multipath effects, surrounding noises, and many other factors impact the quality of the communication link. Test results from tests performed in a particular area may not align with the same tests performed in another area with a different approach. It is also essential to analyze as many configurations as possible to explore the effects of changing physical parameters on the link quality. This paper analyzes the link quality in terms of RSSI, SNR, and packet losses by creating numerous configurations and varying the bandwidth, spreading factor, and coding rate. The experiment's setting is the Kavrepalanchowk district of Nepal, which can be considered a rural area with extremely hilly terrain and minimum surrounding noises that may affect the LoRa signals. This paper not only emphasizes performance evaluation but also demonstrates a simple, low-cost, and reliable LoRaWAN implementation in such an area utilizing the time-division multiple access technique. Moreover, this paper demonstrates the possibility of further integration of such a system with cloud-based platforms for IoT applications.

## 3. Methods

The methodology involves the design of the nodes' hardware architecture, their development, testing in outdoor scenarios, and analysis of key performance metrics varying LoRa parameters such as spreading factor, bandwidth, and coding rate. To demonstrate the IoT application of such a system, the data was pushed in the ThinkSpeak cloud platform from the receiving node after reception and presented in graphical form. Additionally, a Time Division Multiple Access (TDMA) approach was employed to enable efficient resource sharing among multiple nodes operating within a single communication channel.

### 3.1. Sensor Node

The sensor node comprises three main components: an ultrasonic sensor (HC-SR04) as the source of original data, an Arduino Uno as the microcontroller, and the LoRa SX1278 module as the transceiver, as shown in Figure 1. Two LoRa sensor nodes of similar architecture were developed. The power supply to both the sensor and the transceiver was from the 3.3V pin of the microcontroller, which itself was powered by an external battery. Arduino IDE environment was used to program the microcontrollers [14].

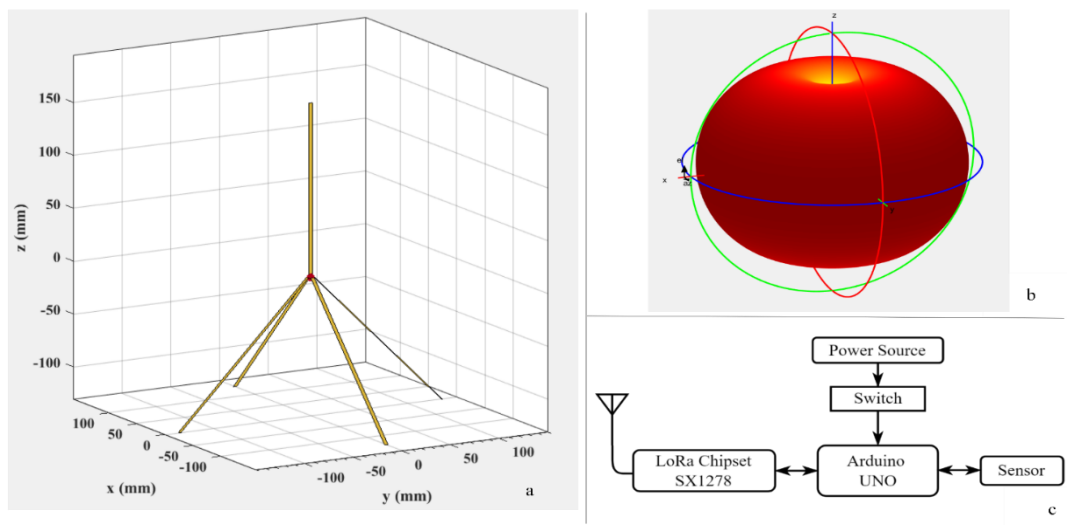

Figure 1: Sensor Node. (a) Antenna structure with radiating element and the ground radials. (b) The radiation pattern of the antenna in 3D. (c) Block diagram of the transmitting node.

The physical structure of the used antenna simulated in MATLAB and its ideal radiation pattern in 3D are shown in Figure 1a and Figure 1b, respectively. The antenna used in this study was a self-made quarter-wave monopole [15] constructed to operate at 433 MHz. The antenna's dimensions were carefully calculated to ensure optimal performance at the desired frequency. The radiating element of the antenna, made from scrap copper wire, was 16.5 cm long, which corresponds to a quarter of the wavelength (λ/4) at 433 MHz.

Additionally, the antenna included ground radials, each 18.4 cm long (slightly longer than λ/4), angled at 45 degrees concerning the plane. The radiation pattern of a quarter-wave monopole antenna with a 4-radials ground plane is omnidirectional, resembling a doughnut shape in the horizontal plane. This means the antenna radiates (and receives) signals equally well in all horizontal directions but with reduced effectiveness directly above and below it. The four radial elements of the ground plane play a crucial role, acting as reflectors to create this pattern and stabilize the antenna's performance [30]. Regarding the antenna's gain, the theoretical approach suggests that a quarter-wave monopole antenna exhibits a gain of approximately 5.15 dBi [31]. This estimation derives from the fact that a half-wave dipole antenna typically has a gain of 2.15 dBi, and the monopole configuration effectively doubles this due to its ground plane reflection, adding an additional 3 dB [30].

### 3.2. Receiving Node

Unlike the sensor nodes, the receiving node, also called the gateway, comprises a LoRa SX1278 transceiver, an ESP32 microcontroller, and a reliable power source.

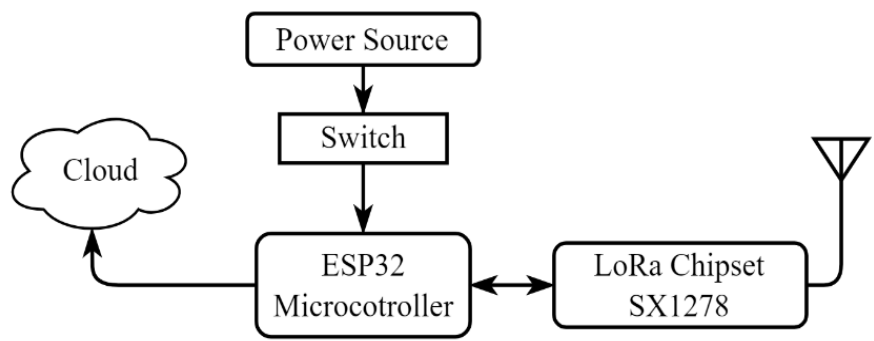

Figure 2: Block diagram of the receiving node.

Figure 2 represents the block diagram of the receiving node. The same type of quarter-wave monopole antenna used in the sensor node was used with the transceiver. The selection of ESP32 as the microcontroller enabled us to program a mechanism to push the received data directly into the cloud platform utilizing the inbuilt WiFi module [16]. The ESP32 board was also programmed using Arduino IDE software.

### 3.3. Parameter Selection

In any communication link, it is crucial to select appropriate parameters to optimize the link's performance. The parameters selected with their different values which are altered during the experiment, are shown in Table 1.

Table 1: LoRa parameters with their adjustable values.

| Parameter | Values |
|---|---|
| Bandwidth (in kHz) | 10.4, 20.8, 62.5, 125, 250, 500 |
| Spreading Factor (chips/symbol) | $2^7, 2^8, 2^9, 2^{10}, 2^{11}, 2^{12}$ |
| Coding Rate | 4/8, 5/8, 6/8, 7/8 |

These parameters are critical in maximizing link quality and energy efficiency in long-range, low-power wireless technologies like LoRa [8]. Each parameter value has trade-offs, and this selection depends on the terrain conditions and the urban/rural context of the link.

Bandwidth is the range of frequencies that can be transmitted over a communication channel. The bandwidth affects the data rate and the receiver's sensitivity [18]. The spreading factor is a parameter in LoRa technology that determines a symbol's duration and the signal's spreading over a more comprehensive frequency range. The symbol duration significantly affects the sensitivity of the receiver [19]. The coding rate is a ratio of information bits to the total number of bits transmitted. It is a parameter that can be adjusted in LoRa technology to improve the link quality [20].

### 3.4. Performance Metrics Selection

To evaluate the impact of physical layer parameters on the performance of a LoRa-based communication system, link quality-defining metrics such as SNR, RSSI, and packet loss were selected. The following subsections briefly explain these metrics and their overall impact on a communication system.

#### 3.4.1. Received Signal Strength Indicator (RSSI)

RSSI indicates the power level of the received signal and serves as an essential indicator of signal quality. It is given by the formula:

$$RSSI = 10 \log_{10}(\frac{P_{received}}{P_{reference}}) \quad (1)$$

The SX1278 chipset has a built-in receiver that continuously samples the incoming signal. It uses an Analog-to-Digital Converter (ADC) to convert these samples into a digital form. The chipset then calculates the RSSI based on these digital samples, often averaging multiple samples to obtain a more accurate estimate. The RSSI value is typically expressed in dBm and represents the strength of the received signal, including both the desired signal and any background noise [21]. In the experiment, the 'packetRssi' function of the 'LoRa.h' library reads the raw RSSI value from the 'RegPktRssiValue' register of the SX1278 chipset [28]. This value is further converted into dBm using the formula:

$$RSSI = RSSI_{from\ the\ register} - Offset \quad (2)$$

where Offset is a specific value determined by the chip's configuration. The Offset value of the chip operating at 433 MHz is 157 dB [28]. Higher RSSI values indicate more robust and reliable signal reception [20].

#### 3.4.2. Packet Loss

Packet loss occurs when the transmitted packet fails to reach its destination due to channel impairments causing loss of information. The total number of packets sent was recorded using a counter in the program embedded in the sending node. The total number of packets received was also measured, which gave insights about the number of packets lost. Then, the packet loss was calculated using the formula:

$$Packet\ Loss = \frac{Numbers\ of\ packets\ lost}{Total\ numbers\ of\ packets\ sent} \times 100 \qquad (3)$$

The physical layer parameters influence packet loss in LoRa communication. Lower spreading factors and narrower bandwidths result in more frequent packet collisions and, consequently, higher packet losses [22].

### 3.4.3. Signal-to-Noise Ratio (SNR)

SNR plays a vital role in wireless connections, particularly in long-range communication, as it can be used to determine the quality of the received signal. It is the ratio of the power of the received signal to the power of the noise components [23, 30, 31]. In wireless communication, the SNR can be calculated using the following formula:

$$SNR = 10\ log_{10}(\frac{P_{signal}}{P_{noise}}) \qquad (4)$$

where $P_{signal}$ is the power of the signal and $P_{noise}$ is the power of the background noise.

The SX1278 chipset calculates the SNR by first estimating the noise level in the channel. This is often done by measuring the power in a part of the spectrum where no signal is present, such as by averaging multiple samples [28]. The chipset then measures the power of the received LoRa signal and calculates the SNR as the difference between the signal power and the estimated noise level. In the experiment, the 'packetSnr' function from the 'LoRa.h' library reads this raw SNR value from the "RegPktSnrValue" register and it is often scaled to fit into a specific data format, such as a signed 8-bit value. SNR affects the Received Signal Strength Indicator (RSSI) and is correlated with the SF and BW of the signal used [24]. A higher SNR implies better signal quality and lower noise [25].

### 3.4.4. Path Loss (PL) Calculation in a LoRa Network

Path Loss signifies the reduction of signal power as it propagates through space. It is a critical factor affecting the performance of LoRa communication systems. This section discusses the process involved in path loss calculation for the developed LoRa network. The SNR and RSSI are initially obtained to determine the Effective Signal Power (ESP). The ESP is a crucial step in calculating path loss, as it accounts for the power of the received signal after considering noise interference and is calculated using the formula:

$$ESP = RSSI + SNR - 10log_{10}(1 + 10^{0.1SNR}) \qquad (5)$$

The actual path loss of the network can be calculated utilizing ESP along with the transmission power of the LoRa SX1278 chipset and the antenna gains. The antenna gains were calculated theoretically according to their types and construction [32]. The empirical formula for calculating the path loss in LoRa networks is:

$$PL = Pt + Gt + Gr - ESP \qquad (6)$$

where *'PL'* is path loss, *'Pt'* is the transmitted power, *'Gt'* is the gain of the transmitter's antenna and *'Gr'* is the receiver's gain. Subsequently, an ideal theoretical baseline for the loss in space unobstructed by any objects and not affected by the multipath effect can be assessed by calculating Free Space Loss (FSL), derived from the Friis Transmission Equation [7], which is given by:

$$FSL = 20log_{10}(d) + 20log_{10}(f) - 20log_{10}(c) + 20log_{10}(4\pi) \qquad (7)$$

where *'d'* is the distance between the transmitter and the receiver, *'f'* is the operating frequency and *'c'* is the speed of the electromagnetic waves. A significant difference between the FSL and actual path loss was noticed, resulting from the presence and influence of factors like losses from other RF components and environmental conditions. This difference was calculated for all the tested configurations and plotted against the physical-layer parameters to analyze their dependencies and relation. Although SF and BW are not directly related to path loss, they influence the ESP, affecting the path loss calculations. This analysis is discussed in detail in the Results section.

**3.5. Experimental Setup for Performance Evaluation**

The performance evaluation was conducted using an experimental setup consisting of a transmitting node and a receiving node installed at a remote site in the Kavrepalanchowk district of Nepal, characterized by its hilly and mountainous terrain. This experimental setting presents a unique environment for studying the performance of the LoRa technology. The district's topography includes varied elevations and undulating landscapes, with rugged hills and valleys, as shown in Figure 3, making communication difficult, specifically during disasters and calamities.

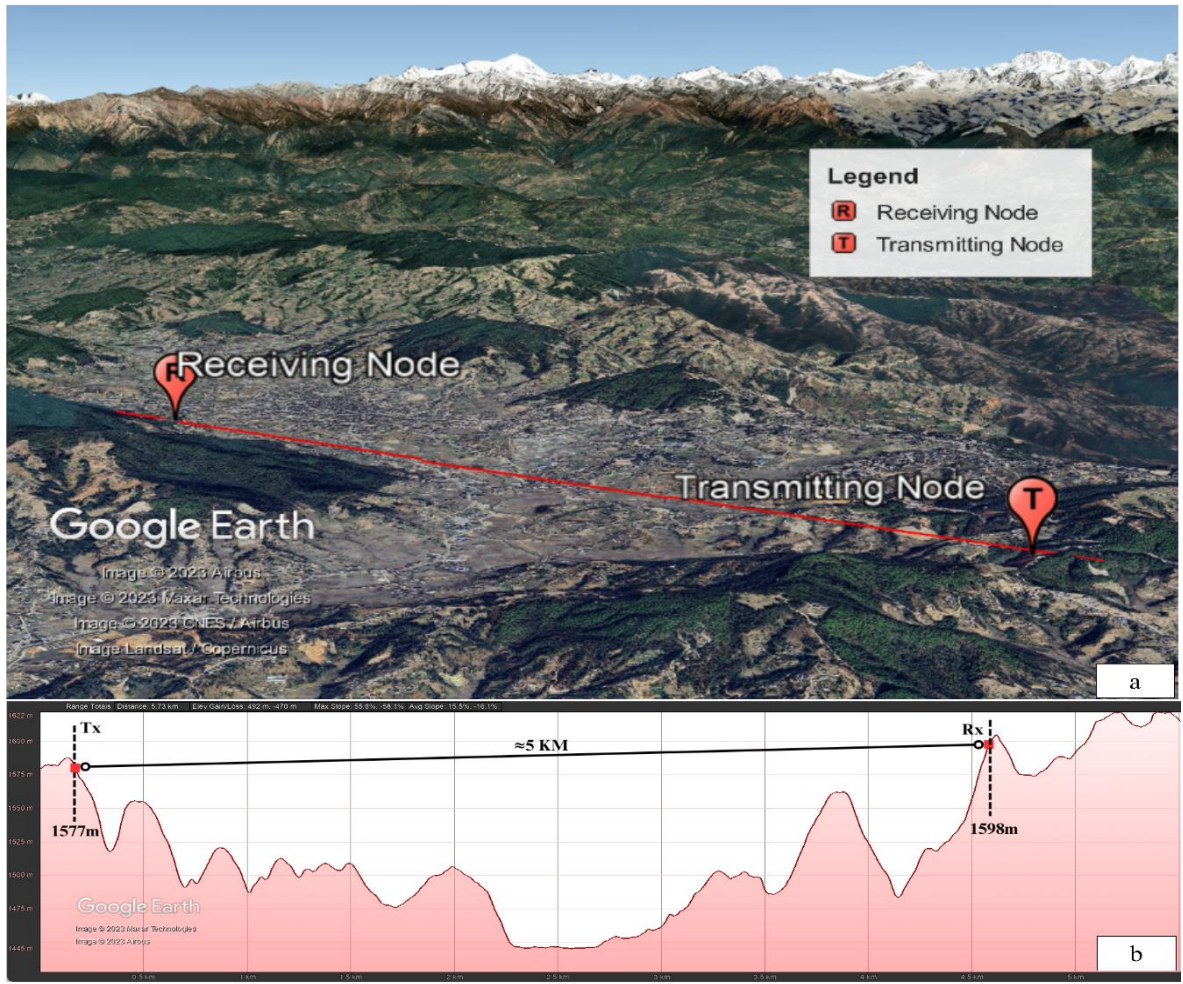

Figure 3: Geographical setting of the experimental area. (a) Terrain map view. (b) Elevation profile of the experimental area.

The region's elevation varies dynamically, with heights ranging from about 1,000 to over 3,000 meters above sea level [33]. Temperate forest makes up the majority of the area's moderate to sparse vegetation [34]. The absence of significant urban development or industrial infrastructure minimizes the risk of man-made electromagnetic interference, making it an ideal setting for testing the efficacy of LoRa technology in a predominantly natural environment. The district's remote setting, with scattered settlements and limited infrastructural developments, underscores the need for reliable, long-range communication technologies like LoRa, especially for applications in remote monitoring and data collection in agricultural or environmental contexts. The existence of high hills and rugged terrain can be major issues affecting the wireless propagation of the signals. In our study, the node placements were executed in such a way that there existed a clear line of sight between the transmitting node and the receiving node. As shown by the elevation profile in Figure 3b, the transmitting node and the receiving node were at the altitudes of 1598m and 1577m above sea level, respectively. The altitude of the valley, which can be considered as the ground base, was approximately 1450m on average. Both of the antennas were clamped to a 1m tall rod and the point-to-point distance between them was approximately 5 kilometers.

The site had an average temperature of 23.5°C and 86% humidity for the entire test duration. The transmitting node utilized a LoRa SX1278 transceiver with an Arduino microcontroller, while the receiving node employed the same transceiver with an ESP32 microcontroller. Quarter-wave monopole antennas, designed in-house, were used at both ends, operating at a frequency of 433 MHz. Performance metrics such as SNR, RSSI, and packet loss were assessed for various combinations of spreading factors, bandwidths, and coding rates. The packets' payload size varied from 1 byte to 2 bytes. These data were the random distance readings per second from the ultrasonic sensor (HC-SR04) [26] and can be considered a low-data-rate requiring system. RSSI and SNR were calculated using the built-in functions of the 'LoRa.h' [27] library, whereas the number of packets being sent and received were recorded over a timeframe to calculate the packet loss. The data retrieved from these setups was fed into MATLAB for further analysis and graphical representations that helped to fulfill the purpose of this experiment.

### 3.6. LPWAN Configuration

To establish a simple LPWAN configuration with limited resources, two sensor nodes with Arduino microcontrollers and LoRa SX1278 transceivers were deployed, along with a gateway consisting of an ESP32 microcontroller and a LoRa SX1278 transceiver.

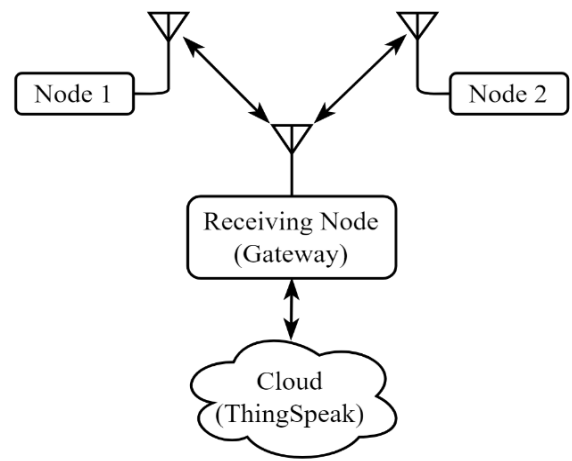

Figure 4: Block diagram of the LPWAN configuration.

Figure 4 shows the block diagram of a multi-node and single-gateway system with IoT integration. The ESP32 microcontroller had an embedded WiFi module for Internet connectivity. TDMA was employed to share a single channel in the time domain. For this, each node was assigned a unique sync word, a 4-digit hexagonal number. The gateway utilized such a sync word for node identification and allocated time slots for data transmission from the sensor nodes to the gateway. The experiment used analog data from an ultrasonic sensor (HC-SR04) as the payload [26]. The connection was terminated after the time slot duration, and the gateway attempted to establish a new connection with the other sensor node. The received packet at the gateway was parsed, manipulated as required, and then pushed to the ThingSpeak cloud platform for further analysis.

# 4. Results and Discussion

After the successful deployment of the nodes, performance evaluation tests were performed to evaluate the system's feasibility. SNR, RSSI, and packet loss are the link quality-defining metrics taken into account through this test. Similarly, another test was successfully conducted to examine the simple LoRaWAN configuration utilizing TDMA in order to establish communication between multiple nodes with the available resources. In addition to these evaluations, further tests were conducted to assess the network's robustness, scalability, and adaptability to different environmental conditions. These assessments provided valuable insights into the network's real-world applicability and potential challenges. Finally, the sensor data was successfully uploaded into a cloud platform to demonstrate the IoT application of such a network in a particular scenario, and the results were plotted. These results are discussed in detail in the following sections.

## 4.1. RSSI Test

Table 2: RSSI Values for Combinations of SF and BW.

| SF/BW | 7      | 8      | 9      | 10     | 11      | 12     |
|-------|--------|--------|--------|--------|---------|--------|
| 10.4  | -92.8  | -93.8  | -110   | -110.8 | -111.2  | -111.9 |
| 20.8  | -91    | -92.66 | -109.8 | -110.6 | -109.4  | -109   |
| 62.5  | -89.2  | -91.8  | -108.4 | -109.6 | -108.25 | -108.2 |
| 125   | -87.28 | -91.83 | -108   | -108.6 | -108.6  | -108.8 |
| 250   | -80.5  | -91    | -105.1 | -108.2 | -108    | -108.5 |
| 500   | -81.33 | -90.6  | -103   | -107.8 | -106    | -106.8 |

*Note:* These data are the result of original work from the authors and are publicly available in [29]. Each configuration was assessed through 5 to 7 measurements, and the results were averaged to ensure reliability.

Table 2 illustrates how the spreading factor and bandwidth affect the RSSI. The SF/BW values are listed in the table's first row and first column. The corresponding RSSI values are provided in the table cells. The table shows that the RSSI values are generally low for the higher values of SF, such as SF of 9 to 12 across each BW. This behavior is expected because higher spreading factors result in longer transmission times and increased sensitivity to noise and interference. With a spreading factor of 7 and a bandwidth of 250 kHz the best RSSI of -80.5 was attained.

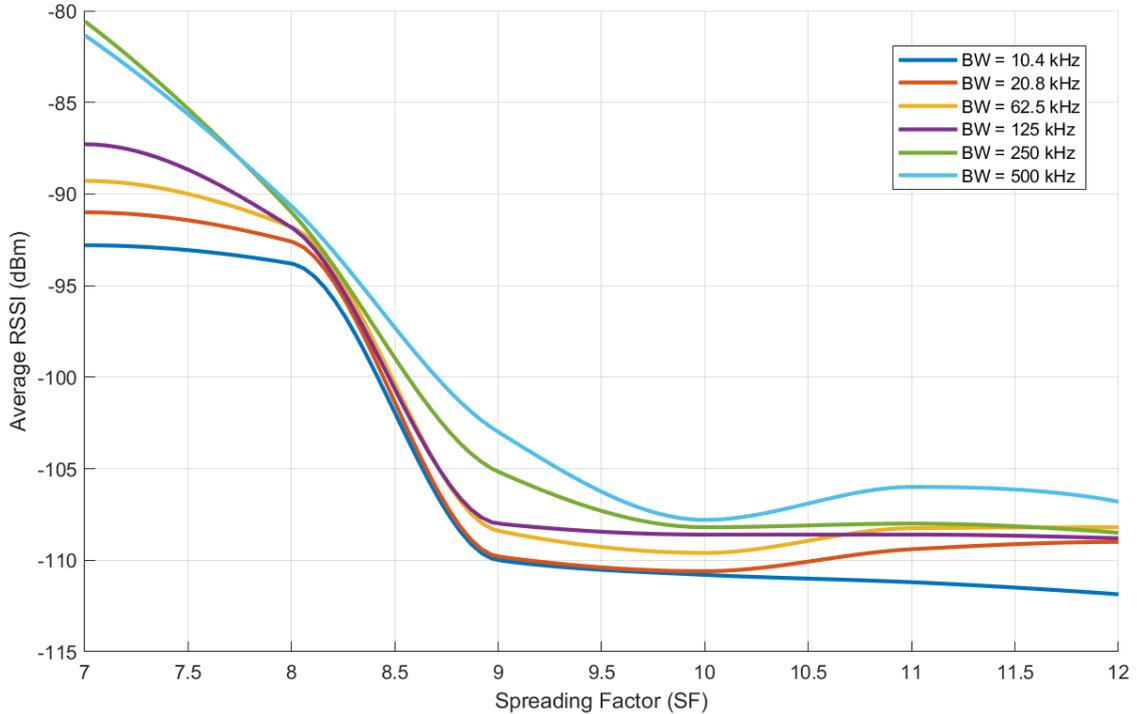

Figure 5: RSSI variation for different bandwidths and spreading factors.

In Figure 5, the x-axis represents the SF, while the y-axis represents the RSSI in dBm. Each line in the plot represents a different bandwidth, with the legend indicating the bandwidth values. We used the Piecewise Cubic Hermite Interpolating Polynomial (PHCIP) interpolation to generate curves for better visualization of the relationship between the SF and the RSSI.

The plot shows that the RSSI value is significantly higher for the lower SF values across all bandwidths, indicating a strong and clear transmission. As the SF increases beyond 8, there is a steep decline in RSSI, indicating a loss in signal quality. This decline stabilizes somewhat from SF 9 to SF 12 across all bandwidths, maintaining a consistently lower RSSI level that does not recover to the heights seen at lower SFs. However, RSSI does not solely depend upon the SF and the BW. Other factors, like specific weather conditions, path loss depending on the experiment's setting, and other external or internal noises, impact the RSSI. It is important to note that the best configuration may vary depending on the specific requirements and constraints of the application. The result suggests that the best RSSI values can be attained with a lower spreading factor for the particular application requiring long-range communication with low data rates, such as data from sensors like an ultrasonic sensor (HC-SR04) in rural areas.

**4.2. SNR Test**

Table 3: SNR Values for Combinations of SF and BW

| SF/BW | 7 | 8 | 9 | 10 | 11 | 12 |
|---|---|---|---|---|---|---|
| 10.4 | 8.4 | 11.55 | 7.85 | 6.05 | 4.5 | 4.3 |
| 20.8 | 8 | 10.25 | 8.8 | 7.6 | 6.6 | 5.55 |
| 62.5 | 8.35 | 10.2 | 7.9 | 9.8 | 7.5 | 6.45 |
| 125 | 8.03 | 10.18 | 8.5 | 7.7 | 8.35 | 6.95 |
| 250 | 7.64 | 10.04 | 7.75 | 7.9 | 7.3 | 6.7 |
| 500 | 3.88 | 5.35 | 5.65 | 6.2 | 5.3 | 4.85 |

*Note:* These data are the result of original work from the authors and are publicly available in [29]. Each configuration was assessed through 5 to 7 measurements, and the results were averaged to ensure reliability.

Table 3 illustrates how the spreading factor and bandwidth affect the SNR. Similar to Table 2, the SF values are listed in the first row, BW values are in the first column, and SNR values are in the respective column.

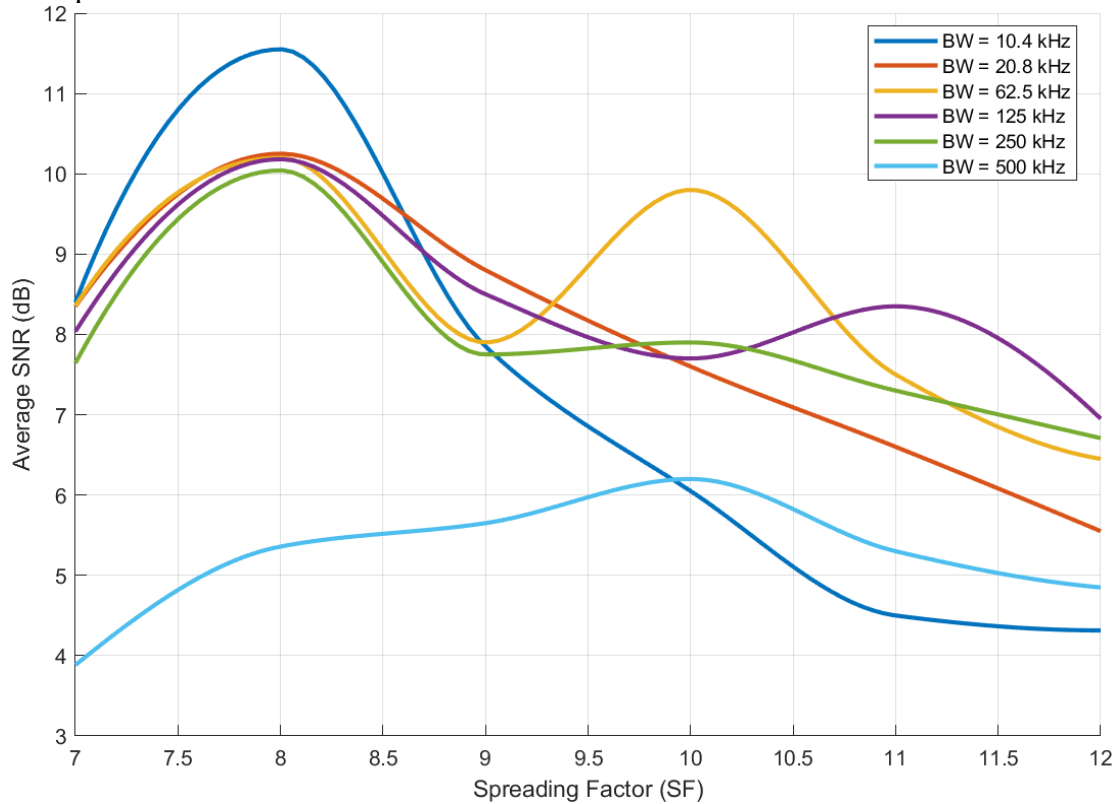

Figure 6: SNR Variation for combinations of bandwidths and spreading factors.

To visualize the data more clearly, Figure 6 illustrates a complex relation between the SNR and the SF for different values of BW. Data from Table 3 were interpolated and plotted using the PCHIP method. The plot indicates that the value of SNR is lower on average for the highest bandwidth, which is 500 kHz. However, for the other bandwidths, the SNR value is better for the lower SF. For the specific application and conditions on which this experiment was conducted, it is clear that the best SNR value was obtained at the configuration where SF was set to 8 and BW was set to 10.4 kHz. It's clear that a lower BW value coupled with a smaller SF will give a better SNR.

Similar to the RSSI, it is essential to note that the SNR value depends on weather conditions, cable or connector losses, path loss, antenna gain, etc. A study focusing on a LoRaWAN deployment in Skellefteå, Sweden, provided insights into how extreme weather conditions, typical of subarctic climates, affect network performance. This research highlighted that colder weather generally improves the SNR. In contrast, warmer conditions tend to compel sensors to select lower SF, minimizing the time on air. The selection of the most optimum values of BW and SF for the best SNR depends upon the application's requirements. However, lower SF and BW will likely result in better SNR for the specific context and settings of our experiment.

## 4.3. Packet Loss Test

Table 1: Packet Loss for Different Values of SF and BW.

| SF/BW | 7      | 8  | 9  | 10 | 11 | 12 |
|-------|--------|----|----|----|----|----|
| 10.4  | 54%    | 0% | 0% | 0% | 0% | 0% |
| 20.8  | 37.50% | 0% | 0% | 0% | 0% | 0% |
| 62.5  | 28.50% | 0% | 0% | 0% | 0% | 0% |
| 125   | 16.60% | 0% | 0% | 0% | 0% | 0% |
| 250   | 0%     | 0% | 0% | 0% | 0% | 0% |
| 500   | 0%     | 0% | 0% | 0% | 0% | 0% |

*Note:* These data are the result of original work from the authors and are publicly available in [29]. Each configuration was assessed through 5 to 7 measurements, and the results were averaged to ensure reliability.

Table 4 represents the packet loss percentages observed during the test for different combinations of SF and BW. The values in each cell indicate the percentage of packet loss observed for the specific SF and BW combination. It is evident from the table that for lower bandwidth values at spreading factor 7, there was significant packet loss. In contrast, no packet loss was detected for any of the other tested configurations, as all the values in the table are recorded as 0%. These findings emphasize the importance of carefully selecting the appropriate SF and BW parameters to mitigate packet loss and ensure reliable data transmission in LoRa-based applications.

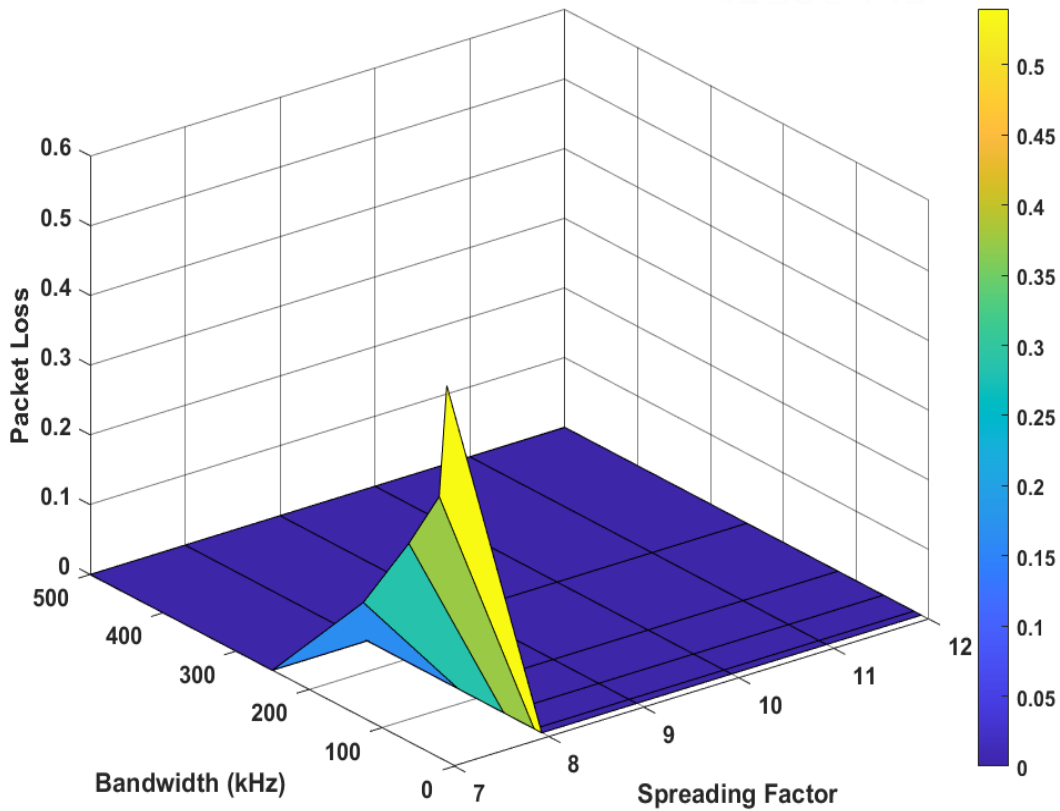

Figure 7: Surface plot of packet loss for multiple combinations of SF and BW.

To identify the patterns and trends from the data, a surface plot shown in Figure 7 provides the three-dimensional approach where the spreading factor is represented on the x-axis, the bandwidth on the y-axis and the packet loss percentage on the z-axis. The colour on the surface plot represents the corresponding packet loss percentage. It can be easily noticed that packet loss is nonexistent in most SF/BW combinations. However, the case is different for narrower bandwidths and the lowest values of the spreading factor. For increasing bandwidth, the packet loss percentage seems to be decreasing. Similarly, for spreading factors greater than 7, the packet loss percentage is null. These results suggest that the choice of physical layer parameters, such as bandwidth and spreading factor, can influence packet loss, indicating certain configurations can be more susceptible to packet loss. By optimizing these parameters, it is indeed possible to reduce the issue of packet loss and, hence, improve the system's performance.

### 4.4. Losses in LoRa Network

In communication systems, various forms of losses such as connector losses, cable losses, and path losses, are pivotal in determining the overall signal strength and system performance. Connector losses are attributed to the discontinuities at the connector interfaces, whereas cable losses result in signal attenuation as it travels through the cable. Both are critical but often overlooked factors. In our setup, the cable losses were deemed negligible due to the use of high-quality 1.13mm SMA Female to U.FL connectors and the short cable length, coupled with the low insertion loss (<0.15dB) of the SMA connectors. In our setup, careful attention was paid to ensure that the antenna's construction and placement maximized its potential gain and maintained the necessary impedance matching with the rest of the system. This practice of impedance matching tends to minimize the losses due to reflection. While the antenna's actual performance was not directly measured due to resource constraints, the system's overall performance and the stability of the received signal strengths suggest effective antenna operation within the expected parameters.

The path loss was calculated with two different approaches, initially utilizing the empirical formula for calculating path loss in LoRa networks and by using another approach that adapted the calculation of FSL considering an ideal scenario without any obstructions and multipath reflection effect. Although the cable losses and connector losses in our setup were insignificant, there was still a significant difference between FSL and actual path loss. This demonstrated the effect of environmental and other external factors on the network.

Table 5: Path Loss difference for different values of SF and BW

| SF/BW | 7 | 8 | 9 | 10 | 11 | 12 |
|---|---|---|---|---|---|---|
| 10.4 | 24.532 | 25.239 | 41.805 | 42.909 | 43.664 | 44.371 |
| 20.8 | 22.738 | 24.137 | 41.483 | 42.441 | 41.405 | 41.213 |
| 62.5 | 21.023 | 23.342 | 40.198 | 41.178 | 40.106 | 40.232 |
| 125 | 19.065 | 23.377 | 39.719 | 40.427 | 40.338 | 40.744 |
| 250 | 12.407 | 22.556 | 36.986 | 39.998 | 39.887 | 40.486 |
| 500 | 13.968 | 22.922 | 35.191 | 39.879 | 38.269 | 39.175 |

*Note:* These data are the result of original work from the authors and are publicly available in [29]. Each configuration was assessed through 5 to 7 measurements, and the results were averaged to ensure reliability.

Table 5 demonstrates the differences between FSL and actual PL for multiple configurations of SF and BW. Changes in SF and BW can vary the network's susceptibility to noise and cause altercations in the RSSI value. The depreciation of signal strength is related to path loss; hence, tuning these parameters will also impact path loss. The severity and trends of path loss for various configurations can be realized in the surface plot shown in Figure 9.

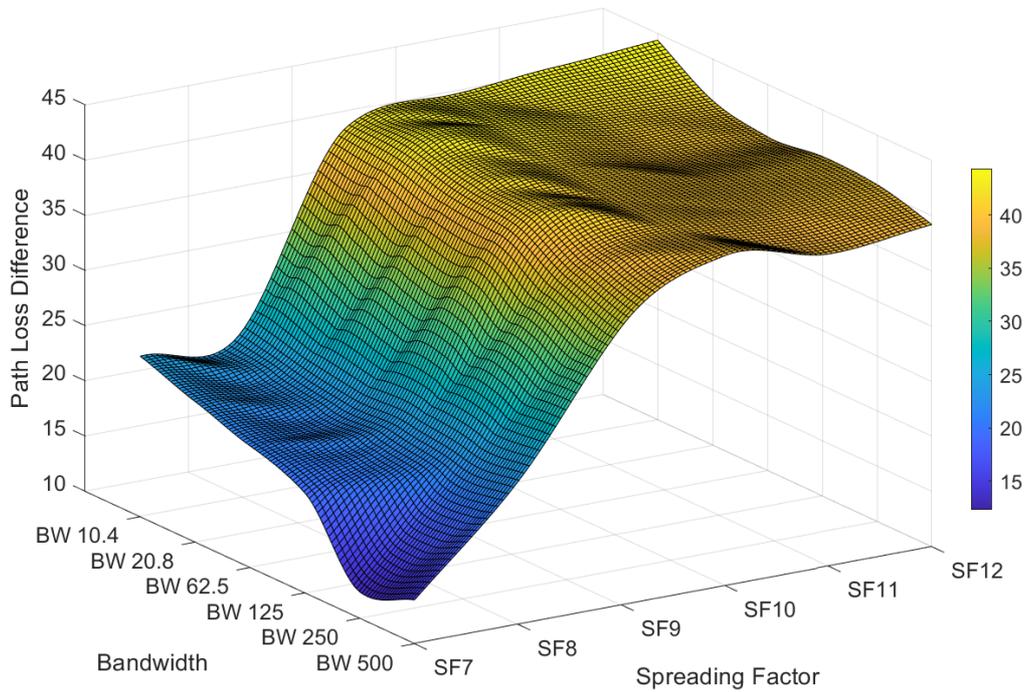

Figure 8: Surface plot of the difference in path loss for multiple combinations of SFs and BWs.

The three axes in Figure 8 include a range of SF from 7 to 12, bandwidth from 10.4 kHz to 500 kHz, and the differences between the obtained path losses. For instance, a general increase in path loss was noted as the SF increased from SF7 to SF12. Interestingly, the higher bandwidths demonstrated the lowest path loss difference, which means the path loss is comparatively closer to the theoretical value of FSL. The data reveals that a configuration with a higher bandwidth value from 125kHz to 250kHz at a lower SF value of 7 and 8 exhibits the lowest path loss progression among the tested parameters, suggesting an optimal balance between signal spread and environmental interference.

In our study, while we did not directly measure the varying impact of weather conditions, it is important to consider the findings of Luomala and Hakala, who conducted an extensive analysis of the effects of temperature and humidity on radio signal strength in outdoor wireless sensor networks [35]. Their research found that signal strength is significantly influenced by ambient temperature and humidity. Specifically, they observed a negative linear relationship between temperature and signal strength, where an increase in temperature generally led to a decrease in signal strength. Furthermore, they noted that high relative humidity could also impact signal strength, especially in temperatures below $0°$ C. Although our study didn't directly measure the impact of varying weather conditions, the consistent weather during our test period (moderate temperature and humidity with minimal cloud cover) likely contributed to the stability of our results. Furthermore, changes in weather conditions can impact the SNR, RSSI, packet loss, and path loss. As this research discussed the dependencies of these performance metrics on the physical-layer parameters in the most common meteorological condition of the region, there is a possibility of considering the obtained relationships as the baseline performance for further finetuning in a different scenario.

## 4.5. Optimal selection of SF and BW

We analyzed the available data across several metrics to determine the most effective LoRa configuration for the specific environments. The higher bandwidths, such as 125 and 250 kHz, configured with the lower SF values, show relatively good RSSI and lower path loss. However, the SNR at these bandwidths is not consistently high, indicating potential issues with signal clarity. SF 8 across three bandwidths (62.5 kHz, 125 kHz, and 250 kHz) shows improved SNR values. This consistency in SNR at SF 8 makes it the preferred choice for the spreading factor. Given the trade-offs in selecting bandwidths, 125 kHz and 250 kHz could be considered if the deployment scenario can tolerate a bit less SNR. BW of 62.5 kHz offers a better SNR, which is advantageous for maintaining clear communications over longer distances, making it suitable for expansive rural areas.

Although all three bandwidths (62.5 kHz, 125kHz, 250kHz) can be chosen, for the specific geographic terrain with such high hills and mountains and applications requiring long-range communication with a low data rate, the optimum configuration of LoRa's physical parameters would be a bandwidth of 62.5 kHz and SF 8.

## 4.6. Effect of Coding Rate on SNR

Table 6: SNR Values for Different Coding Rates

| CR/BW | 4/8 | 5/8 | 6/8 | 7/8 |
|---|---|---|---|---|
| 250 | 9.75 | 8.88 | 8 | 5.65 |

*Note:* These data are the result of original work from the authors and are publicly available in [29]. Each configuration was assessed through 5 to 7 measurements, and the results were averaged to ensure reliability.

Considering the balanced SNR at the bandwidth of 250 kHz and SF of 8, Table 4 illustrates the effect of the coding rate on the SNR. In this table, different values of coding rates are listed in the first row, and the bandwidth is listed in the first column. The obtained SNR values are listed in the corresponding cells. It can be observed that as the coding rate increases, the SNR values decrease. This indicates that the lower coding rates provide a better SNR, allowing robustness against noise and interference. During the experiment, the best value of SNR was obtained at a coding rate of 4/8.

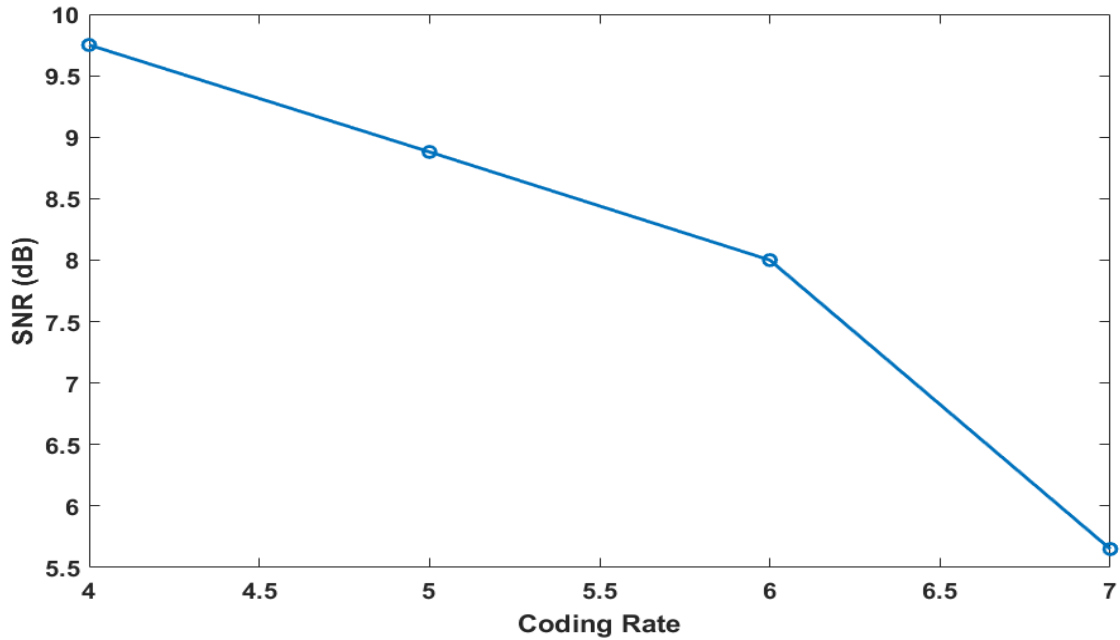

Figure 9: SNR variation for different coding rates.

Figure 9 shows a graphical representation of the experiment's output plotted to visualize the effect of the coding rate on the signal-to-noise ratio (SNR). A higher coding rate can increase susceptibility to noises and interferences in the LoRa-based communication system. The observed relationship between coding rate and SNR indicates the need for careful consideration and optimization of coding rate selection to achieve the desired performance for specific applications.

**4.7. TDMA and IoT Implementation**

A simple TDMA technique was employed to acquire bandwidth sharing between two nodes in the time domain. This setup successfully allocated the channel dynamically to different nodes in their respective time slots. The gateway node first authenticated and established a connection with the sensor nodes utilizing a synchronization hex-word of 4 bytes. Following the connection establishment, the sensor node sends the data, an ultrasonic sensor reading, to the gateway node. This newly received packet on the receiving side was parsed and pushed to a cloud platform using its API key. Meanwhile, the next time slot was utilized to communicate and follow the same process with the other node.

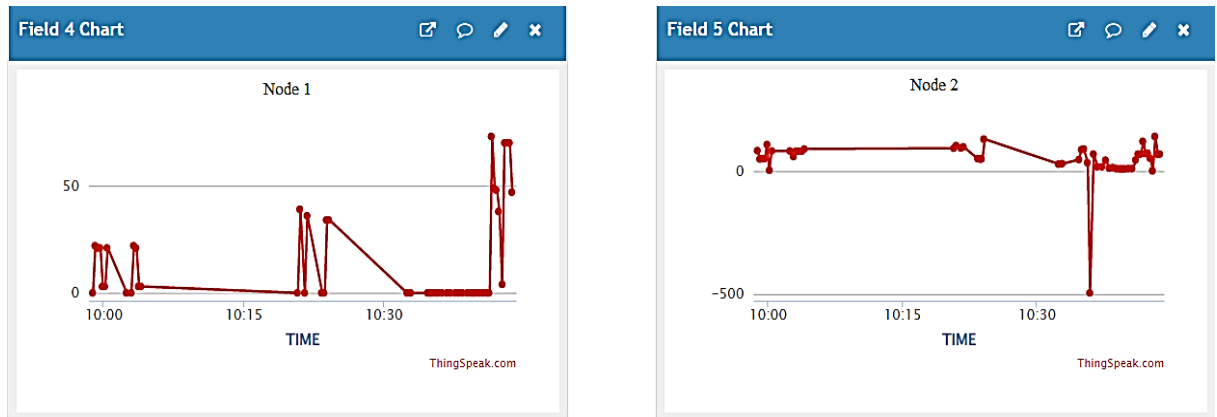

Figure 10: Data from multiple nodes being plotted in the cloud.

In our study, the distance data from the ultrasonic sensors were parsed and pushed to the cloud. Figure 10 demonstrates the data reception from two sensor nodes to the ThingSpeak cloud. The x-axis and y-axis show timestamps of data reception and the magnitude of the random noisy data (distance in cm) from the ultrasonic sensor, respectively. This demonstrates that the developed framework can be further developed and applied in an IoT-based system in a geographically challenging scenario.

## 5. Conclusion

The scope of LoRa technology is immense and holds great promise for addressing the unique challenges and opportunities of rural settings, specifically in the context of Nepal. This research paper is focused on evaluating the impact of physical-layer parameters on the network performance and sheds light on the optimum LoRa configuration for wireless communication between nodes. Analyzing metrics such as RSSI, SNR, and packet loss revealed the importance of carefully selecting parameters such as bandwidth, spreading factor, and coding rate. Also, the research involved the design and development of a simple LoRaWAN network comprising two sensor nodes and one receiving node integrated with the LoRa SX1278 transceiver. ESP32 was employed at the receiving node, allowing direct IoT implementation in the network. The experimental results suggest that the RSSI values are significantly lower for the increasing SF across all the bandwidths. For lower spreading factors and narrow bandwidths, the SNR value is higher. Also, setting up a lower coding rate will contribute to a higher SNR value. Additionally, packet loss was significantly higher for the narrower bandwidth and lower value of the spreading factor. The path loss analysis demonstrated that the signal's susceptibility towards external factors is less in higher bandwidths and a lower spread factor. For the specific geographic terrain considered, with such high hills and mountains and applications requiring long-range communication with low data rate, the optimum configuration of LoRa's physical parameters would be a bandwidth of 62.5 kHz, SF 8 and a coding rate of 4/8. This configuration offers a good balance between signal strength (RSSI), signal quality (SNR), and reliability (Packet Loss), ensuring a robust and efficient data transmission. However, while designing such systems, the trade-off with other parameters like data rate, bit error rate, latency, etc, must be considered according to the application's requirements, which becomes a limitation of this study. It can also be concluded that the TDMA approach can be deployed in such environments to enhance communication efficiency and scalability. The precise synchronization and channel separation introduced by TDMA significantly reduced the interferences in the network and facilitated the effective utilization of the available bandwidth. The results also demonstrate the compatibility of such systems with cloud-based platforms like ThingSpeak, enabling the creation of a cost-effective and remotely accessible IoT system for real-time monitoring and analysis. In the context of challenging geographical settings like in the experimental scenario, the investigated technology, when configured as recommended, can provide long-range communication suitable for low-data rate applications with minimal infrastructure. This includes Wireless Sensor Networks (WSNs) for disaster management, continuous monitoring in advanced agriculture, and climate monitoring activities. In essence, this research illustrates the versatility of LoRa technology in Nepal's distinct topography, offering a robust solution for remote connectivity challenges.

## 6. Conflicts of Interest

On behalf of all authors, the corresponding author states that there is no conflict of interest.

## 7. Funding Statement

This work has been done by a group of a professor and his students and it was not funded by any financial bodies.

## 8. Supplementary Materials

The collected data, calculations, and code files used for programming the microcontrollers during the experiment along with the MATLAB codes used for data visualization can be found on a public database [29, 36].